\documentclass[twocolumn,amsmath,amssymb,prc,aps]{revtex4-1}
\usepackage{graphicx}
\usepackage{color}
\usepackage{bm}
\usepackage{hyperref}

\usepackage{scalerel}
\usepackage{tikz}
\usetikzlibrary{svg.path}
\definecolor{orcidlogocol}{HTML}{A6CE39}
\tikzset{
  orcidlogo/.pic={
    \fill[orcidlogocol] svg{M256,128c0,70.7-57.3,128-128,128C57.3,256,0,198.7,0,128C0,57.3,57.3,0,128,0C198.7,0,256,57.3,256,128z};
    \fill[white] svg{M86.3,186.2H70.9V79.1h15.4v48.4V186.2z}
                 svg{M108.9,79.1h41.6c39.6,0,57,28.3,57,53.6c0,27.5-21.5,53.6-56.8,53.6h-41.8V79.1z M124.3,172.4h24.5c34.9,0,42.9-26.5,42.9-39.7c0-21.5-13.7-39.7-43.7-39.7h-23.7V172.4z}
                 svg{M88.7,56.8c0,5.5-4.5,10.1-10.1,10.1c-5.6,0-10.1-4.6-10.1-10.1c0-5.6,4.5-10.1,10.1-10.1C84.2,46.7,88.7,51.3,88.7,56.8z};
  }
}

\newcommand\orcidicon[1]{\href{https://orcid.org/#1}{\mbox{\scalerel*{
\begin{tikzpicture}[yscale=-1,transform shape]
\pic{orcidlogo};
\end{tikzpicture}
}{|}}}}

\begin{document}


\title{
  Shape transition of Nd and Sm isotopes
  and 
  neutrinoless double-beta decay nuclear matrix element of $^{150}$Nd
}

\author{Yusuke Tsunoda \orcidicon{0000-0002-0398-1639}}
\affiliation{Center for Computational Sciences,
  University of Tsukuba, 1-1-1 Tennodai, Tsukuba 305-8577, Japan}
  
\author{Noritaka Shimizu \orcidicon{0000-0003-1638-1066}}
\email{shimizu@nucl.ph.tsukuba.ac.jp}
\affiliation{Center for Computational Sciences,
  University of Tsukuba, 1-1-1 Tennodai, Tsukuba 305-8577, Japan}
\affiliation{Center for Nuclear Study, The University of Tokyo, 7-3-1 Hongo, Tokyo 113-0033, Japan}
\author{Takaharu Otsuka \orcidicon{0000-0002-1593-5322}}
\affiliation{Department of Physics, The University of Tokyo, 7-3-1 Hongo, Tokyo 113-0033, Japan}
\affiliation{RIKEN Nishina Center, 2-1 Hirosawa, Wako, Saitama 351-0198, Japan}
\affiliation{Advanced Science Research Center, Japan Atomic Energy Agency, Tokai, Ibaraki 319-1195, Japan}

\begin{abstract}
  Neutron-rich Nd and Sm isotopes are known to exhibit shape phase transition as a function of neutron number. Among them, $^{150}$Nd and $^{150}$Sm are important not only because they are transitional nuclei, 
  but also the parent and daughter nuclei of double-beta decay.
  We performed large-scale shell-model calculations of 
  even-even Nd and Sm isotopes 
  including the spherical-deformed shape transition. 
  The quasi-particle vacua shell model enables 
  us to perform shell-model calculations with sufficiently 
  large model space with the $^{110}$Zr inert core.
  The shell-model result well reproduces 
  the experimental excitation energies 
  and quadrupole properties of the yrast and non-yrast states.
  The nuclear matrix element of neutrinoless double-beta decay of $^{150}$Nd 
  is evaluated showing its modest enhancement by shape mixing. 
\end{abstract}

\maketitle 


Atomic nuclei show various shapes of their surfaces as the proton number ($Z$) and the neutron number ($N$) change, for example, from a sphere to a weakly distorted ellipsoid to a strongly distorted one.   
The Nd and Sm isotopes of $N \geq 82$ are one of the clean examples of this change, or evolution, from spherical shapes to ellipsoidal deformed shapes \cite{ringshuck}.  
While this shape evolution is an interesting and important subject, systematic studies covering both spherical and deformed cases on equal footing must handle two different many-body structures, spherical and deformed, which correspond to significantly different types of mean-field solutions \cite{ringshuck}.  Besides this difficulty, the situations between these two limits, usually called transitional, must be described equally well, which likely requires more sophisticated multi-nucleon treatments.  

The Interacting Boson Model (IBM) \cite{iachello_arima_book} provides the eigensolution of its Hamiltonian within empirical approaches, and was applied to the shape evolution of Sm isotopes \cite{scholten_1978}, nicely describing it as a U(5)-SU(3) transition. 
A plenty of theoretical works have been devoted to understand this evolution 
\cite{Cejnar_RevModPhys.82.2155,Nomura_PhysRevLett.101.142501,Robledo_shape_PhysRevC.78.034314,Li_PhysRevC.79.054301,Delaroche-PhysRevC.81.014303}.
Besides this, the shell model is, in principle, an ideal approach, because it solves the multi-nucleon 
Schr\"odinger equation with a given effective nucleon-nucleon ($NN$) interaction, without referring to the specific features ({\it e.g.} shape) of the solutions to be obtained.
On the other side, the practical application of the conventional shell-model calculations is limited by the number of valence nucleons and/or the number of valence single-particle orbits, which can result in exploding dimension of the Hamiltonian matrix to be diagonalized.  This limitation emerges for the study of the Nd-Sm shape evolution.  Quite recently, however, a breakthrough was made by the Monte Carlo Shell Model (MCSM) (see reviews \cite{ppnp_mcsm,mcsm_ptep,phys_scr_mcsm}) as a different formulation of the shell-model calculation, and the description of the shape evolution in Sm isotopes was given in the light of realistic $NN$ interactions \cite{SmEr_PRL_2019}.

In recent years, neutrinoless double-beta decay attracts keen and broad interests, as it is a crucial key to elucidate whether the neutrino is a Majorana particle or not \cite{review0nbb}.
The double-beta decay from $^{150}$Nd to $^{150}$Sm may be used to extract the nuclear matrix element (NME) of the neutrinoless double-beta ($0\nu\beta\beta$) decay, which is sensitive to the structure of involved nuclei \cite{Lopez-PRL111}. This article presents a new value of the NME of this decay. 

Many experiments to search for neutrinoless double-beta decay have been planned and undergone.  
Each of them employs one of a dozen of the nuclides which can undergo double-beta decay and provides us with the lower limit of the half-life for $0\nu\beta\beta$ decay.
To derive the information of neutrino mass from the half-life, the NME value for each nuclide has to be estimated theoretically. However, this estimation brings about large uncertainties of factor 2 \cite{engel_menendez}. 
Currently, the smallest upper limit of the effective neutrino mass reaches around 0.1 eV by the experiment with $^{136}$Xe \cite{Gando_PRL}.
If we restrict ourselves to the case of $^{150}$Nd, there were quite a few efforts to evaluate its NME, such as the IBM \cite{Barea_IBM_PhysRevC.79.044301}, the quasiparticle random-phase approximation (QRPA) \cite{FangFaessler2010}, 
and the generator coordinate method (GCM) 
\cite{song_nme,Yao_150Nd_octupole,rodriguez_EDF_150Nd,yao_ppnp,Lopez-PRL111}.
Another GCM study employing a relativistic density functional revealed that the contribution of the octupole correlations cancels 7\% of the NME \cite{Yao_150Nd_octupole}.
The recent study of the projected shell model showed the importance of triaxial deformation \cite{triaxialpsm}.
Thus, various many-body correlations indeed need to be treated for the precise estimation of the NME. 

The earlier MCSM calculation successfully describes the yrast states of Sm isotopes \cite{SmEr_PRL_2019}. However, the precise evaluation of the NME of the $0\nu\beta\beta$ decay requires an improved efficiency towards more precise treatment of pairing correlations.
In the present work, the quasi-particle vacua shell model (QVSM) \cite{qvsm} is adopted for this purpose, as its high efficiency towards high precision really pays for heavier computer resources needed.  
We made a test calculation of the present NME in Ref.~\cite{qvsm} using the QVSM, and found, in terms of moment of inertia, that the model space had to be enlarged. 
In the present study, we enlarge the model space by including the breaking of the $^{132}$Sn core so that the experimental moment of inertia of the Nd and Sm isotopes can be reproduced well,  and evaluate the $0\nu\beta\beta$-decay NME of $^{150}$Nd.


In the QVSM framework, a shell-model eigen wave function is described as a linear combination of 
the angular-momentum-, parity-, and number-projected quasiparticle vacua  \cite{ringshuck} as
\begin{equation}
  \label{eq:qvsmwf}
  |\Psi_{N_b} \rangle = \sum_{n=1}^{N_b}\sum_{K=-J}^{J}
  f^{(N_b)}_{nK} P^{J\pi}_{MK} 
  P^Z  P^N | \phi_n \rangle,
\end{equation}
where $P^Z$, $P^N$, and $P^{J\pi}_{MK}$ denote, respectively,  
the projectors of the proton-number, the neutron-number, 
and the angular-momentum and parity combined.  Here, 
$|\phi_n\rangle = | \phi_n^{(\pi)} \rangle \otimes | \phi_n^{(\nu)} \rangle $
is a product of the quasiparticle vacua of protons and neutrons \cite{ringshuck}. 
The coefficient, $f^{(N_b)}_{nK}$, is an amplitude in the linear combination, 
and its value is determined by solving the generalized eigenvalue problem:
\begin{eqnarray}
  \label{eq:geneq}
  && \sum_{n=1}^{N_b} \sum_{K=-J}^J \langle \phi_m|HP^{J}_{MK}P^Z P^N 
  |\phi_n \rangle f^{(N_b)}_{nK}
  \nonumber \\
  &=& E^{(N_b)} 
      \sum_{n=1}^{N_b}\sum_{K=-J}^J
      \langle \phi_m|P^{J}_{MK}P^Z P^N |\phi_n \rangle f^{(N_b)}_{nK},
\end{eqnarray}
with $E^{(N_b)}$ being the energy eigenvalue. 
The quasi-particle vacua $|\phi_n\rangle$ are determined so that $E^{(N_b)}$ is minimized. 
In the case of $N_b=1$, it corresponds to the Hartree-Fock-Bogoliubov (HFB) calculation with the variation after the angular-momentum, parity and number projection.
In that sense, the QVSM is an extension of the variation after projection with the superposition.
It is stressed that each QVSM basis vector can carry pairing correlations in it to good extents, but the pairing correlations are incorporated mainly through superpositions among different basis vectors in the MCSM (by Slater determinants).
As the present NME has features common with pairing correlations, the QVSM is expected to yield NME values more precise for a given number of basis vectors than the MCSM, meaning a faster convergence.  
It was shown in~\cite{qvsm} that the $0\nu\beta\beta$-decay NME converged smoothly even with a small number of the QVSM basis vectors.

The model space and the interaction are taken from our previous MCSM study \cite{SmEr_PRL_2019} with a minor modification.
The model space consists of the $sdg$ shell, $0h_{11/2}$, $1f_{7/2}$, and $2p_{3/2}$ orbits for protons 
and $pfh$ shell,
$0i_{13/2}$, $1g_{9/2}$, $2d_{5/2}$, and $3s_{1/2}$ orbits for neutrons
so that it contains $\Delta j=\Delta l = 2$ pairs in respective upper shells.
The Hamiltonian is constructed by combining the monopole-based universal ($V_\textrm{MU}$) interaction \cite{vmu} whose proton-neutron interaction is multiplied by a factor of 0.94 to its isoscalar central part (as in \cite{SmEr_PRL_2019}), and the $G$-matrix-based interaction for the proton-proton and neutron-neutron interactions \cite{brown2000}. 
The proton (neutron) pairing matrix elements are multiplied by a factor of 0.9 (0.7) 
and some of the single-particle energies are slightly tuned from Ref.~\cite{SmEr_PRL_2019}. 
The pairing correlations are handled more efficiently by the QVSM compared to the MCSM with Slater determinants.  For heavy nuclei, previous MCSM calculations may have led to stronger pairing interactions, because of a fit with a finite number of the basis vectors.  Thus, the weaker pairing interactions suggested by the QVSM are natural consequences for heavy nuclei.  
The contamination of the center-of-mass excitation 
is removed by adding the Lawson term with 
$\beta_{CM}\hbar\omega/A =1.0$ MeV resulting 
its quanta $O(10^{-3})$, which is sufficiently small. 

In the present calculation for the $84\le N\le 92$ nuclei, 24 QVSM basis states are optimized so as to lower the two lowest energy eigenvalues of $0^+$, $2^+$, and $4^+$.
The many-body subspace is then spanned by these 72 basis states and the eigenvalue problem in Eq.(\ref{eq:geneq}) is solved.
For $N=82$ and 94, 16 QVSM basis states are used to optimize the lowest energy eigenvalues of $0^+$, $2^+$ and $4^+$, and totally 48 basis states are used to make up the QVSM wave function.
The effective charges are taken as $(e_p,e_n)=(1.6,0.6)e$ throughout the present work.

\begin{figure}[thbp]
  \includegraphics[scale=0.4]{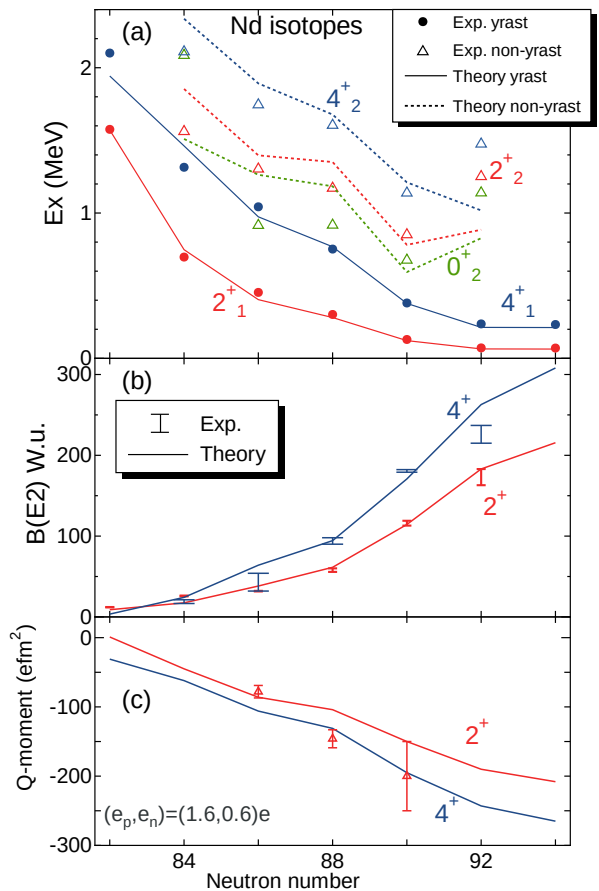}
  \caption{
    (a) Excitation energies, 
    (b) $B(E2;2^+_1 \rightarrow 0^+_1)$ and $B(E2;4^+_1 \rightarrow 2^+_1)$ values,
    and (c) spectroscopic quadrupole moments of the $2^+_1$ and $4^+_1$ states of Nd isotopes
    against the neutron number. 
    (a) The red and blue solid lines (filled circles) denote  
    the theoretical (experimental) values of the $2^+_1$ and $4^+_1$ states, respectively. 
    The green, red, and blue dotted lines (open triangles) denote those of the $0^+_2$, 
    $2^+_2$, and $4^+_2$ states, respectively. 
    (b,c) Solid lines are theoretical values, while symbols denote experimental values.
  }
  \label{fig:Nd-ex-E2}
\end{figure}
Figure \ref{fig:Nd-ex-E2} shows the evolution of the excitation energies, the E2 transition probabilities, and the spectroscopic quadrupole moments of Nd isotopes. 
At the $N=82$ semimagic nucleus, the $2^+$ and $4^+$ excitation energies are rather large, and the seniority scheme is expected to work. With increasing neutron number, the excitation energies gradually decrease, and the ratio of the $4^+_1$ and $2^+_1$ energies indicates the transition from the spherical vibrator to the rotational band. 
The gradual increase of the E2 transition probabilities and quadrupole moments support this interpretation.
The theoretical excitation energies of the non-yrast states of $^{152}$Nd ($N=92$) are a few hundred keV lower than the experimental values, suggesting possible unmeasured levels.

\begin{figure}[thbp]
  \includegraphics[scale=0.4]{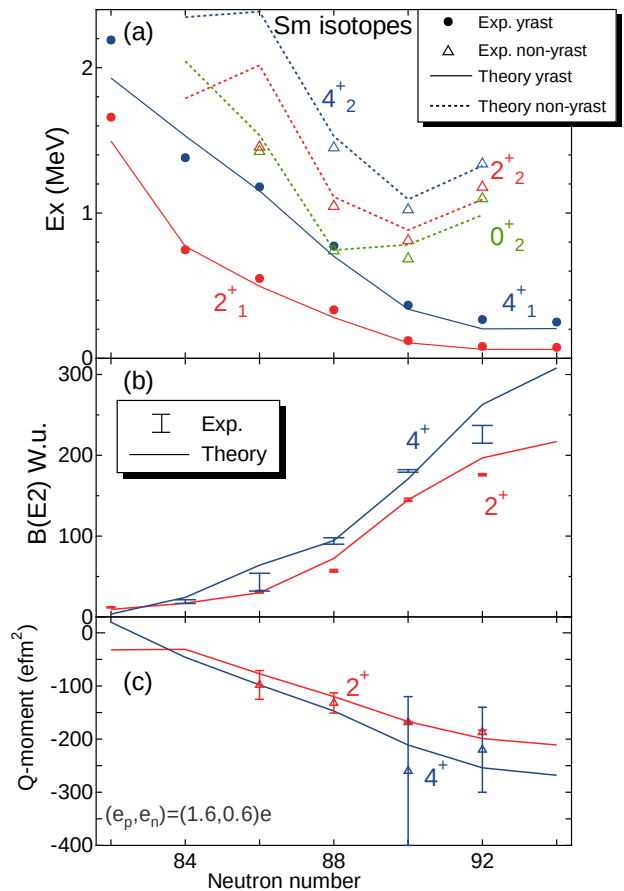}
  \caption{
    (a) Excitation energies, 
    (b) B(E2) values, 
    and (c) spectroscopic quadrupole moments of Sm isotopes. 
    See the caption of Fig.~\ref{fig:Nd-ex-E2}.
  }
  \label{fig:Sm-ex-E2}
\end{figure}

Figure \ref{fig:Sm-ex-E2} shows the excitation energies, $B(E2)$ values, and spectroscopic quadrupole moments of Sm isotopes.
The $2^+_1$ and $4^+_1$ excitation energies gradually decrease, and the quadrupole collectivity grows as $N$ increases.  The Nd isotopes show similar trends.
The most remarkable difference between Nd and Sm isotopes is seen at the non-yrast states: 
the excitation energies of the $0^+_2$, $2^+_2$, and $4^+_2$ states drop down abruptly between $N=86$ and 88 for Sm isotopes, whereas such change occurs between $N=88$ and 90 in the Nd isotopes. This tendency is nicely reproduced in the present results.

\begin{figure*}[htbp]
  \includegraphics[scale=0.5]{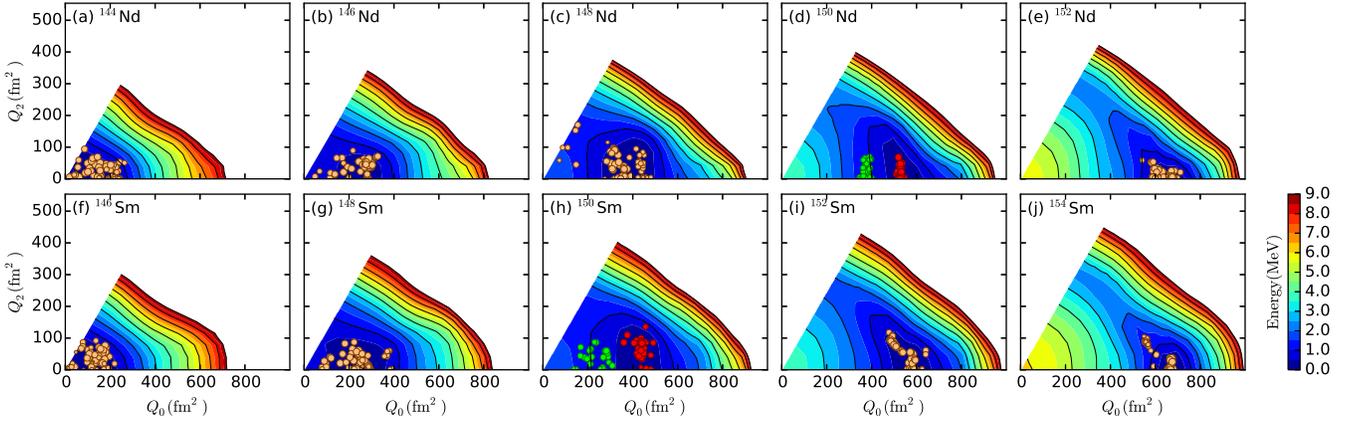}
  \caption{
    T-plots of the $0^+_1$ states of 
    $^{144,146,148,150,152}$Nd and $^{146,148,150,152,154}$Sm by the QVSM wave functions.
    On the figures of $^{150}$Nd and $^{150}$Sm, the green and red circles denote the ``S" and ``L" groups, respectively. See the main text for details.
  }
  \label{fig:tplot}
\end{figure*}

The QVSM wave function can be analyzed by using the T-plot figures, 
in which its component is visualized as scattered circles on the energy surface in the same way as the MCSM  \cite{Ni_PRC_YT,jpg_tplot}. 
Figure~\ref{fig:tplot} shows the T-plots of the $0^+_1$ states of the Nd and Sm isotopes.
The ground-state energy is drawn as contour lines given by the constrained number-projected 
Hartree-Fock-Bogoliubov (HFB) calculations with the constraints given by usual intrinsic quadrupole moments, $Q_0 = \langle \phi | \hat{Q}_{20} P^ZP^N|\phi \rangle$ and $Q_2 = \langle \phi|\hat{Q}_{22}P^ZP^N|\phi\rangle$, where $\hat{Q}_{20}$ and $\hat{Q}_{22}$ are the quadrupole operators \cite{ringshuck}.
The shell-model Hamiltonian is taken for this calculation.
The positions and areas of the circles on the energy surface represent, respectively, the deformation and the overlap probabilities 
between $|\phi_i\rangle$ and $|\Psi_{N_b} \rangle$ in Eq.~(\ref{eq:qvsmwf}).
At $N=84$ and 86, the minimum of the energy surfaces is close to the 
$\langle Q_0\rangle=\langle Q_2\rangle=0$ 
and their T-plot points are concentrated around the spherical shape. 
As $N$ increases, the minimum of the energy surface and T-plot distributions gradually move toward prolate deformation.
Among them, the T-plots of $^{150}$Nd and $^{150}$Sm show a characteristic feature: the distribution is divided into two groups. 
We focus on their structures since these nuclides are the parent and daughter nuclei of the double-beta decay.

\begin{figure}[htbp]
  \includegraphics[width=0.9\linewidth]{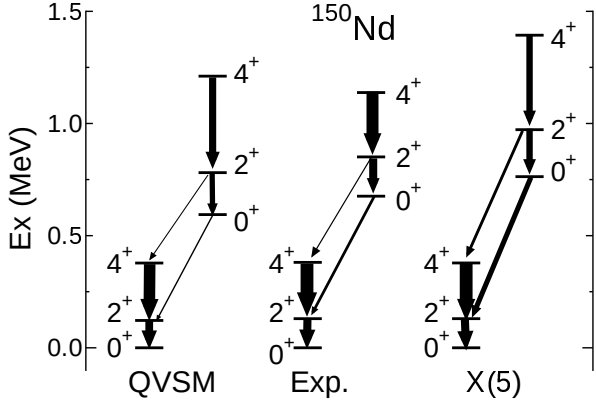}
  \caption{
   Partial level schemes of $^{150}$Nd obtained by the QVSM, the experiment, and the X(5) critical point symmetry \cite{X5_Iachello, krucken_X5_150Nd}.
   The arrow widths are proportional to the $B(E2)$ values. Only those $>$ 10 W.u. are displayed.
   The X(5) results are scaled so that it reproduces the experimental values of  Ex$(2^+_1)$ and $B(E2;2^+_1\rightarrow 0^+_1)$.
  }
  \label{fig:Nd150level}
\end{figure}

Figure~\ref{fig:Nd150level} shows the partial level scheme of the QVSM in comparison with the experimental ones.   
The nucleus $^{150}$Nd has been considered a candidate for the critical point symmetry X(5) \cite{X5_Iachello,krucken_X5_150Nd}. This model assumes that the energy surface is flat in the direction of $\beta$ on the prolate side, while the present work gives a shallow minimum by $\sim$ 4 MeV at the prolate deformation 
(see Fig.~\ref{fig:tplot}(d)).
Figure~\ref{fig:Nd150level} includes the result of the X(5) model. These three results agree with each other reasonably well, up to $B(E2)$ values. 

The NME of $0\nu\beta\beta$ decay is evaluated with the closure approximation as 
\begin{equation}
  M^{0\nu}=\langle 0^+_f|\hat{O}|0^+_i\rangle
  =M^{0\nu}_{GT}-\frac{g_V^2}{g_A^2}M^{0\nu}_F+M^{0\nu}_T,
\end{equation}
where $GT$, $F$ and $T$ denote, respectively, the contributions of Gamow-Teller type, Fermi type and tensor type \cite{senkovhoroi}.
Here, $|0^+_i\rangle$ and $|0^+_f\rangle$ are ground-state wave functions of $^{150}$Nd and $^{150}$Sm, respectively.
The intermediate energy of the this approximation is taken from the empirical formula, $E_c=1.12 A^{1/2}$ MeV \cite{ec_empirical} and $g_A/g_V$=1.27 is adopted.
The two-body current contribution to the transition operator is not included.
\begin{figure}[htbp]
  \includegraphics[width=\linewidth]{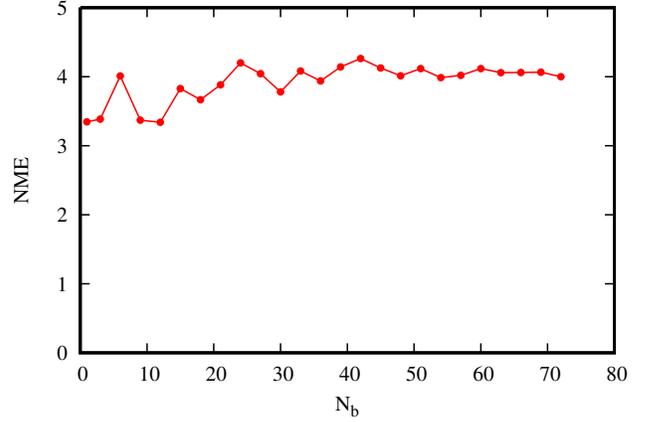}
  \caption{
    $0\nu\beta\beta$-decay NME of $^{150}$Nd without short-range correlation, $M^{0\nu}$, 
    against the number of QVSM basis vectors, $N_b$.
  }
  \label{fig:nme}
\end{figure}
Figure~\ref{fig:nme} displays the convergence of the $0\nu\beta\beta$-decay NME of $^{150}$Nd as a function of the number of the QVSM basis vectors. It shows rather good convergence at $N_b=72$.
The present result for $N_b=1$ is similar to but beyond the HFB calculation, because the variation after the angular-momentum, parity, and number projections is performed. 
The contribution of the effect of configuration mixing in the QVSM framework increases the NME by 20\% in comparison with the value of $N_b=1$.
The QVSM basis vectors are obtained so that the two low-lying energies are minimized. Even if three low-lying energies are optimized, the resultant NME shows a difference of less than 5\%. 
The NME calculated without short-range correlation becomes 4.00 as shown in Table \ref{tab:nme}. 
In addition, we applied three types of short-range correlations (SRCs): Miller-Spencer \cite{millerspencer}, 
CD-Bonn and Argonne \cite{SimkovicSRC}.
The Miller-Spencer short-range correlation quenches the NME by $\sim$20\%, and the CD-Bonn and Argonne correlations give 10\% uncertainty.
The tensor contributions are negligibly small like other theoretical studies.

\begin{table}[htbp]
  \centering
  \begin{tabular}{ccccc}
    \hline 
    \hline 
    SRC    & $M^{0\nu}_{GT}$ & $M^{0\nu}_{F}$ &$M^{0\nu}_{T}$ &$M^{0\nu}$ \\
    \hline 
    None   &  3.32        &  -1.11        & -0.01  &  4.00 \\
    Miller-Spencer &  2.65&  -0.91        & -0.01  &  3.21 \\
    CD-Bonn& 3.40         &  -1.16        & -0.01  &  4.11 \\
    Argonne& 3.21         &  -1.10        & -0.01  &  3.88 \\
    \hline 
    \hline 
  \end{tabular}
  \caption{Nuclear matrix elements of neutrinoless double-beta decay with various short-range correlations.}
  \label{tab:nme}
\end{table}

In other theoretical methods, \textit{e.g.}, the quasi-particle random phase approximation gives NME=2.71 \cite{QRPA_CH}. 
The NME values given by the latest generator-coordinate method based on the relativistic energy density functional are 5.6 \cite{Yao_PRC91}
and 5.2 \cite{Yao_150Nd_octupole}.
It was suggested that the large 
difference between the initial and final state deformations would suppress the NME \cite{song_nme,rodriguez_EDF_150Nd,FangFaessler2010}.

Since the decay life of $2\nu\beta\beta$ decay of $^{150}$Nd to the $0^+_2$ state of $^{150}$Sm has been experimentally measured \cite{polischuk2021double}, the $0\nu\beta\beta$ NME to the $0^+_2$ state might be possible and worth mentioning.
The $0\nu\beta\beta$-decay NME from the ground state of $^{150}$Nd to the $0^+_2$ state of $^{150}$Sm is 1.1, which is much smaller than that of the ground state like the result of the relative energy density functional \cite{song_nme}.

\begin{figure}[htbp]
  \includegraphics[scale=0.45]{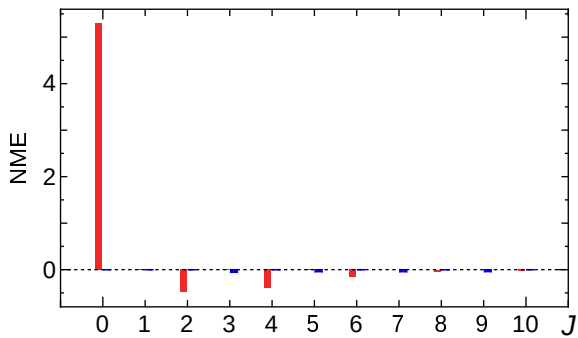}
  \caption{
    $J^\pi$-decomposition of the $0\nu\beta\beta$-decay NME of $^{150}$Nd. 
    The red (blue) bars denote positive (negative) parity components.
  }
  \label{fig:nme-j}
\end{figure}
Figure \ref{fig:nme-j} shows the decomposition of the angular momentum $J$ and the parity $\pi$ of the NME of $^{150}$Nd. 
The decomposition of $J^\pi$ intermediate states is defined in the same way as in Refs.~\cite{iwata_nme,giuliani2012neutrinoless} as 
\begin{eqnarray}
  M^{0\nu} 
  &=& \sum_{J,\pi} 
      \langle 0^+_f| \sum_{i\le j, k\le l} M^{J,\pi}_{i,j,k,l}
      [ [c^\dagger_i \otimes c^\dagger_j ]^{(J,\pi)}
    \nonumber \\
    && [c_k\otimes c_l]^{(J,\pi)} ]^{(0)}
      | 0^+_i \rangle .
\end{eqnarray}
The contribution of the $0^+$ pair plays a dominant role, while the $2^+$ and $4^+$ pairs have a certain negative contribution.
In previous shell-model studies of the $0\nu\beta\beta$ decays of $^{48}$Ca \cite{iwata_nme}, $^{82}$Se, and $^{130}$Te \cite{giuliani2012neutrinoless}, the contribution of the $2^+$ pair is rather large (around $-2$) and cancels most of the contribution of the $0^+$ pair. In the present case, the contribution of the $2^+$ pair is rather small and its cancellation is not strong. 
The contribution of the intermediate negative-parity states is negligible.

Now we discuss the nuclear structures of $^{150}$Nd and $^{150}$Sm, utilizing more details of the T-plots. The T-plots in Fig.~\ref{fig:tplot} show that the points are concentrated in two regions: the two groups of different deformations form the ground-state wave functions, where one group is for weaker deformation (green circles) and the other for stronger deformation (red circles).
The group composed only of smaller (larger) ones is referred to by the index ``S" (``L") hereafter. 
The ground-state wave functions are their linear combinations: 
\begin{eqnarray}
  | 0^+_1; ^{150}\textrm{Nd} \rangle
  &=& 0.50|S; ^{150}\textrm{Nd} \rangle 
  + 0.86|L; ^{150}\textrm{Nd} \rangle
  \nonumber \\
  | 0^+_1; ^{150}\textrm{Sm} \rangle
  &=& 0.65|S; ^{150}\textrm{Sm} \rangle 
  + 0.76|L; ^{150}\textrm{Sm} \rangle .
  \label{eq:lc_LS}
\end{eqnarray}
These two components with different deformations are almost degenerate in energy and are strongly mixed. 
The second $0^+$ states are also described approximately as their linear combinations, 
which are orthogonal to the ground states, and show excitation energies lower than those of the neighboring isotopes. 
The NME 
$\langle S; ^{150}\textrm{Sm} |\hat{O}|S; ^{150}\textrm{Nd} \rangle$ turns out to be 3.87 without the SRC, which is largest among NMEs of other combinations. 
$\langle L; ^{150}\textrm{Sm} |\hat{O}|L; ^{150}\textrm{Nd} \rangle$,
$\langle L; ^{150}\textrm{Sm} |\hat{O}|S; ^{150}\textrm{Nd} \rangle$, 
and $\langle S; ^{150}\textrm{Sm} |\hat{O}|L; ^{150}\textrm{Nd} \rangle$ 
are 2.81, 2.70, and -0.28, respectively.
$\langle S; ^{150}\textrm{Sm} |\hat{O}|L; ^{150}\textrm{Nd} \rangle$ is small because the deformation of the bra and ket states is significantly different.
One sees that the mixing amplitudes in Eq.~(\ref{eq:lc_LS}) modestly enhance the NME to 4.00, and the present calculation appears to treat this feature appropriately.


In summary, we present a shell-model description of the spherical-to-deformed shape evolution of Nd and Sm isotopes including the non-yrast states in a unified way.
The ground states of $^{150}$Nd and $^{150}$Sm are transitional in this shape evolution scenario.  They are characterized by two components of different prolate shapes and are mixed. 
This situation may be regarded to be a ``double-prolate-shape coexistence with mixing". 
The $0\nu\beta\beta$-decay NME of $^{150}$Nd is affected by this mixing and is evaluated as 4.1 with the CD-Bonn short-range correlations. 
A more sophisticated $0\nu\beta\beta$ NME operator is expected to be derived for future studies, for instance, an additional short-range matrix element proposed in Refs.~\cite{cirigliano_new_leading_PhysRevLett.120.202001,jokiniemi2021impact}.

\section*{Acknowledgment}
\label{sec:ack}

The authors acknowledge Javier Men\'{e}ndez and 
 Yutaka Utsuno for valuable discussions. 
This research used computational resources of the supercomputer Fugaku
(hp220174, hp210165, hp200130) at RIKEN Center for Computational Science, Oakforest-PACS supercomputer (Center for Computational Sciences, University of Tsukuba xg18i035), and Wisteria-O supercomputer (Center for Computational Sciences, University of Tsukuba wo22i022). 
This research was supported 
by ``Program for Promoting Researches on the Supercomputer Fugaku''
(JPMXP1020200105) and JICFuS, and the KAKENHI grant (17K05433, 20K03981, 19H05145, 21H00117).

\bibliography{Nd-0nbb}

\end{document}